\RequirePackage{lineno}
\documentclass[aps,twocolumn,superscriptaddress,showpacs,amsmath,amssymb,nofootinbib]{revtex4-1}
 \pdfoutput=1
 \usepackage{graphicx,epsf,amsmath}
 \usepackage[]{latexsym}
 \usepackage{amsmath}
 \usepackage{subcaption}
 \usepackage{float}
 \usepackage{color}
 
 \usepackage{dcolumn}   
 \usepackage{bm}        
 \usepackage{amssymb}   
 \usepackage[utf8]{inputenc}

 \usepackage{subcaption} 
 \usepackage{booktabs}

 \newcommand{\be}{\begin{eqnarray}}
 \newcommand{\ee}{\end{eqnarray}}

\begin{document}

\title{Mass Composition of UHECRs from $X_{\rm max}$ Distributions Recorded by the Pierre Auger and Telescope Array Observatories}

 \author{Nicusor Arsene}
 \email[]{nicusorarsene@spacescience.ro}
 \affiliation{Institute of Space Science, P.O.Box MG-23, Ro 077125 
 Bucharest-Magurele, Romania}

 \date{\today}

\begin{abstract}

In this paper we infer the mass composition of the ultra high energy cosmic rays (UHECRs) from measurements of $X_{\rm max}$ distributions recorded at the Pierre Auger (2014) and Telescope Array (TA) (2016) Observatories, by fitting them with all possible combinations of Monte Carlo (MC) templates from a large set of primary species (p, He, C, N, O, Ne, Si and Fe), as predicted by EPOS-LHC, QGSJETII-04 and Sibyll 2.1 hadronic interaction models. 
We use the individual fractions of nuclei reconstructed from one experiment in each energy interval to build equivalent MC $X_{\rm max}$ distributions, which we compare with the experimental $X_{\rm max}$ distributions of the other experiment, applying different statistical tests of compatibility.
The results obtained from both experiments confirm that the mass composition of the UHECRs is dominated ($\gtrsim$$70\%$) by protons and He nuclei
{in the energy range investigated $\lg E (\rm eV)$ = [17.8--19.3] (Auger) and $\lg E \rm (eV)$ = [18.2--19.0] (TA).} 
The indirect comparisons between the $X_{\rm max}$ distributions recorded by the two experiments show that the degree of compatibility of the two datasets is good, even excellent in some high energy intervals, especially above the ankle ($\lg E (\rm eV) \sim 18.7$). However, our study reveals that, at low energies, further effort in data analysis is required in order to harmonize the results of the two experiments.

\end{abstract}

\maketitle
\section{Introduction}

The UHECRs ($E > 10^{18}$ eV) are the most energetic particles in the universe produced by the most energetic astrophysical objects. Their origin and acceleration mechanisms are still unknown despite the huge efforts in the last decades of the entire astrophysics community \cite{Mollerach:2017idb}. A realistic description of such a complex process of production and transport through the intergalactic medium requires accurate information about the extra-galactic magnetic fields \cite{Harari:2015mal,Mollerach:2020mhr}, interaction with the cosmic microwave background (CMB) and the extra-galactic background light (EBL), distribution and type of sources which accelerate these particles \cite{Thoudam:2016syr, Aloisio:2017qoo, PhysRevD.72.081301, PhysRevD.74.043005, 2012APh....39..129A} and, last but not least, accurate information about their energy spectrum \cite{ABUZAYYAD201593, PhysRevLett.125.121106}, mass composition \cite{PhysRevD.99.022002,Yushkov:2020nhr, Aab_2017, Glushkov:2019oft} and arrival directions \cite{Aab:2017tyv, Aab_2018} measured at ground based cosmic ray experiments. 
These former properties are indirectly reconstructed, making use of different parameters of extensive air showers (EAS) recorded by the cosmic ray experiments such as the Pierre Auger Observatory (Auger) \cite{ThePierreAuger:2015rma} and the Telescope Array (TA) \cite{2012NIMPA.676...54T}.

The mass composition of the primary UHECRs is still a matter of debate even if it was extensively studied by combining different complementary techniques \cite{Aab:2014kda, Abbasi:2018nun, Aab:2014dua, Aab:2016enk, PhysRevD.102.023036}.
The most reliable parameter from the EAS used to infer the mass composition proved to be $X_{\rm max}$, the atmospheric depth where the energy deposit profile of the secondary particles reaches its maximum\cite{GH}. This parameter is related to the primary particle mass as $\langle X_{\rm max}\rangle \propto-\ln A$.

The measurements of mass composition reported by the Auger \cite{Aab:2014kda,Bellido:2017cgf} and TA\cite{Abbasi:2014sfa} experiments, on the basis of the first two moments of the $X_{\rm max}$ distributions ($\langle X_{\rm max}\rangle$ and $\sigma_{X_{\rm max}}$) are not in very good agreement on the entire energy range. While the TA results suggest only a light composition above $E > 10^{18.2}$ eV, the Auger results clearly indicate a transition to a heavier component starting from $E \sim 10^{18.5}$ eV and becoming increasingly heavier at the highest energies.

On the other hand, it was shown that using only the limited information given by the first two moments of a $X_{\rm max}$ distribution 
may lead, in very particular cases, to a misinterpretation of the mass composition, since different mixes of primary particles can reproduce exactly the same $\langle X_{\rm max}\rangle$ and $\sigma_{X_{\rm max}}$ values. To avoid such situations, a method was proposed, which uses the entire shape of each $X_{\rm max}$ distribution by fitting them with Monte Carlo (MC) templates for four fixed primary species (p, He, N and Fe) obtaining in this way information about fractions of individual nuclei \cite{PhysRevD.90.122006}.
Following this approach, the measurements of $X_{\rm max}$ distributions recorded by Auger and TA experiments were indirectly  compared \cite{deSouza:2017wgx} concluding that both measurements are compatible in the limits of statistical and systematical uncertainties.

In a recent paper \cite{Arsene:2020ago}, we showed that by fitting the $X_{\rm max}$ distributions only with four fixed elements (p, He, N and Fe), an artificial worsening of the fit quality can be induced and the reconstructed fractions might be biased as a consequence of a high abundance of some intermediate elements (e.g., Ne/Si) not included in the fitting procedure. 
We argued that a more appropriate approach is to fit the observed $X_{\rm max}$ distributions with all possible combinations of elements from a larger set of primaries (p, He, C, N, O, Ne, Si and Fe) obtaining the ``best combination'' of elements that best describe the data.

In this work, we perform an indirect comparison between $X_{\rm max}$ distributions measured by Auger (2014) \cite{Aab:2014kda} and TA (2016) \cite{Abbasi:2018nun} experiments following the approach proposed in \cite{Arsene:2020ago}. 

In Section \ref{MC}, we describe the simulation procedure for obtaining the MC templates for a large set of primary species (p, He, C, N, O, Ne, Si and Fe), taking into account the detector effects (acceptance and resolution) of both experiments, employing different hadronic interaction models. 
In Section \ref{fit}, we obtain the fractions of individual nuclei which best describe the $X_{\rm max}$ distributions measured by Auger and TA experiments on the entire energy range $\lg E (\rm eV)$ = [17.8--19.3] (Auger) and $\lg E \rm (eV)$ = [18.2--19.0] (TA) following two methods. The first method reconstructs the fractions of individual nuclei of the ``best combination'' of elements which best describe the observed $X_{\rm max}$ distributions. With the second method we extract the average of the reconstructed fractions of each species from all possible combinations of fitting elements whose fit quality was higher than a threshold value, obtaining in this way the evolution of the abundances of  all primary species as a function of energy.  

In Section \ref{compatibility}, we present the compatibility of the measurements of the two experiments on the basis of three statistical tests: \textit{p-value} as goodness-of-fit, Kolmogorov--Smirnov ($KS$) and Anderson--Darling ($AD$). Section \ref{conclusions} concludes the paper.
\section{Monte Carlo templates} \label{MC}

The MC templates ($X_{\rm max}$ distributions or Probability Density Functions (PDFs) of $X_{\rm max}$) for 8 primary species (p, He, C, N, O, Ne, Si and Fe) for each energy interval of $0.1$ in $\lg E (\rm eV)$, employing three hadronic interaction models EPOS-LHC \cite{PhysRevC.74.044902}, QGSJETII-04 \cite{PhysRevD.74.014026} and Sibyll 2.1 \cite{PhysRevD.80.094003} were generated using CONEX v4r37 simulation code \cite{Pierog:2004re, Bergmann:2006yz}. Due to the experimental limitations, the MC templates for the TA experiment were simulated for only eight energy intervals in the range $\lg E (\rm eV)$ = [18.2--19.0] and 15 energy intervals for Auger $\lg E (\rm eV)$ = [17.8--19.3]. The zenith angle of the showers were isotropically sampled in the interval $\theta = [0^{\circ}$--$60^{\circ}]$ { ensuring an isotropic flux on flat surface $dN / dcos(\theta) \sim cos(\theta)$}. The statistics of each $X_{\rm max}$ distribution are of the order of $10^4$--$10^5$ events, with a larger number of simulations for proton induced showers in comparison with the heavier nuclei.

A PDF of $X_{\rm max}$ for a primary nuclear species in a given energy interval consists of a binned $X_{\rm max}$ distribution in the range [0--2000] g/cm$^{2}$ with a bin width $= 20$ g/cm$^{2}$ for the Auger case, while for the TA experiment the $X_{\rm max}$ distributions are binned in the range [500--1300] g/cm$^{2}$ with a bin width $= 40$ g/cm$^{2}$.

To account for detector effects, the true $X_{\rm max}$ values calculated with CONEX were modified in accordance with the acceptance and resolution of each experiment. In the case of Auger, the true $X_{\rm max}$ values were modified by using Equations (7) and (8) from \cite{Aab:2014kda}, while for the TA case the true MC templates were modified in accordance with the biases (reconstruction + acceptance) and resolutions computed in \cite{Abbasi:2018nun} in Table 1 and Figure 9 of that paper, considering full detector simulations for p, He, N and Fe for the QGSJETII-04 model. 
For the intermediate elements, the bias and resolution of $X_{\rm max}$ are obtained using a 2nd degree polynomial interpolation. It is worth mentioning that the possible uncertainties on the bias and resolution of the intermediate elements, artificially introduced by this interpolation, would be much smaller than the experimental resolution of the $X_{\rm max}$ parameter. The exact values of the biases and resolutions used to construct the MC templates for the TA MC templates are listed in Table \ref{bias}. For EPOS-LHC and Sibyll 2.1 models we considered the same values of biases and resolutions as computed for the QGSJETII-04 model. Again, we have to stress that this approximation cannot affect the results significantly, since the possible deviations from the true values of bias and resolution for EPOS-LHC and Sibyll 2.1 cannot exceed more than few g/cm$^{2}$. As stated in \cite{Abbasi:2018nun}, the bias is larger for deeply penetrating showers, for example, protons. As an example, the difference between the average $<X_{\rm max}>$ induced by protons predicted by Sibyll 2.1 and QGSJETII-04 in the energy interval $\lg E (\rm eV)$ = [18.2--18.3] is about $2.4$ g/cm$^{2}$, so the bias of the reconstructed $X_{\rm max}$ for protons predicted by these models should be roughly the same.

\begin{table*}
\centering
\setlength{\tabcolsep}{4pt}

\caption{The values of the biases and resolution in units of [g/cm$^{2}$] used to build MC templates for the case of TA data in each energy interval for 8 primary species.}
\begin{tabular}{lllllllllllllllll} 
\toprule
\multicolumn{1}{c}{} & \multicolumn{2}{c}{p}                                                                           & \multicolumn{2}{c}{He}                                                                         & \multicolumn{2}{c}{C}                                                                          & \multicolumn{2}{c}{N}                                                                          & \multicolumn{2}{c}{O}                                                                          & \multicolumn{2}{c}{Ne}                                                                         & \multicolumn{2}{c}{Si}                                                                         & \multicolumn{2}{c}{Fe}                                                                          \\ 
lgE (eV)             & \begin{tabular}[c]{@{}l@{}}bias \\\end{tabular} & \begin{tabular}[c]{@{}l@{}}res\\\end{tabular} & \begin{tabular}[c]{@{}l@{}}bias\\\end{tabular} & \begin{tabular}[c]{@{}l@{}}res\\\end{tabular} & \begin{tabular}[c]{@{}l@{}}bias\\\end{tabular} & \begin{tabular}[c]{@{}l@{}}res\\\end{tabular} & \begin{tabular}[c]{@{}l@{}}bias\\\end{tabular} & \begin{tabular}[c]{@{}l@{}}res\\\end{tabular} & \begin{tabular}[c]{@{}l@{}}bias\\\end{tabular} & \begin{tabular}[c]{@{}l@{}}res\\\end{tabular} & \begin{tabular}[c]{@{}l@{}}bias\\\end{tabular} & \begin{tabular}[c]{@{}l@{}}res\\\end{tabular} & \begin{tabular}[c]{@{}l@{}}bias\\\end{tabular} & \begin{tabular}[c]{@{}l@{}}res\\\end{tabular} & \begin{tabular}[c]{@{}l@{}}bias\\\end{tabular} & \begin{tabular}[c]{@{}l@{}}res\\\end{tabular}  \\ 
\hline
{[}18.2 - 18.3] & -6.17  & 17.2  & -5.62  & 15.7 & -4.04 & 14.43 & -3.7  & 14.2  & -3.41  & 13.75  & -2.87  & 13.18  & -2.03 & 12.39 & -1.78  & 13.2        \\
{[}18.3 - 18.4] & -6.21  & 17.2  & -5.56  & 15.7 & -4.04 & 14.43 & -3.72 & 14.2  & -3.41  & 13.75  & -2.87  & 13.18  & -2.03 & 12.39 & -1.78  & 13.2        \\                                                                                 
{[}18.4 - 18.5] & -11.07 & 17.2  & -8.65  & 15.7 & -3.27 & 14.43 & -2.17 & 14.2  & -1.17  & 13.75  & -0.55  & 13.18  & -0.81 & 12.39 & -1.54  & 13.2        \\    
{[}18.5 - 18.6] & -9.22  & 17.2  & -7.68  & 15.7 & -4.23 & 14.43 & -3.51 & 14.2  & -2.85  & 13.75  & -1.7   & 13.18  & -0.1  & 12.39 & -1.8  & 13.2        \\                                   
{[}18.6 - 18.7] & -9.31  & 17.2  & -7.13  & 15.7 & -2.33 & 14.43 & -1.37 & 14.2  & -0.49  & 13.75  & -0.99  & 13.18  & -0.83 & 12.39 & -2.34  & 13.2        \\                                 
{[}18.7 - 18.8] & -7.75  & 17.2  & -6.4   & 15.7 & -3.42 & 14.43 & -2.81 & 14.2  & -2.27  & 13.75  & -1.34  & 13.18  & -0.17 & 12.39 & -3.18  & 13.2        \\                                
{[}18.8 - 18.9] & -7.75  & 17.2  & -6.4   & 15.7 & -3.42 & 14.43 & -2.81 & 14.2  & -2.27  & 13.75  & -1.34  & 13.18  & -0.17 & 12.39 & -3.18  & 13.2        \\                                 
{[}18.9 - 19.0] & -7.17  & 17.2  & -6.09  & 15.7 & -3.66 & 14.43 & -3.16 & 14.2  & -2.7   & 13.75  & -1.92  & 13.18  & -0.87 & 12.39 & -2.63  & 13.2        \\                                    
\bottomrule \hline \hline
\end{tabular}
\label{bias}
\end{table*}

\begin{figure}
    \includegraphics[width=0.5\textwidth]{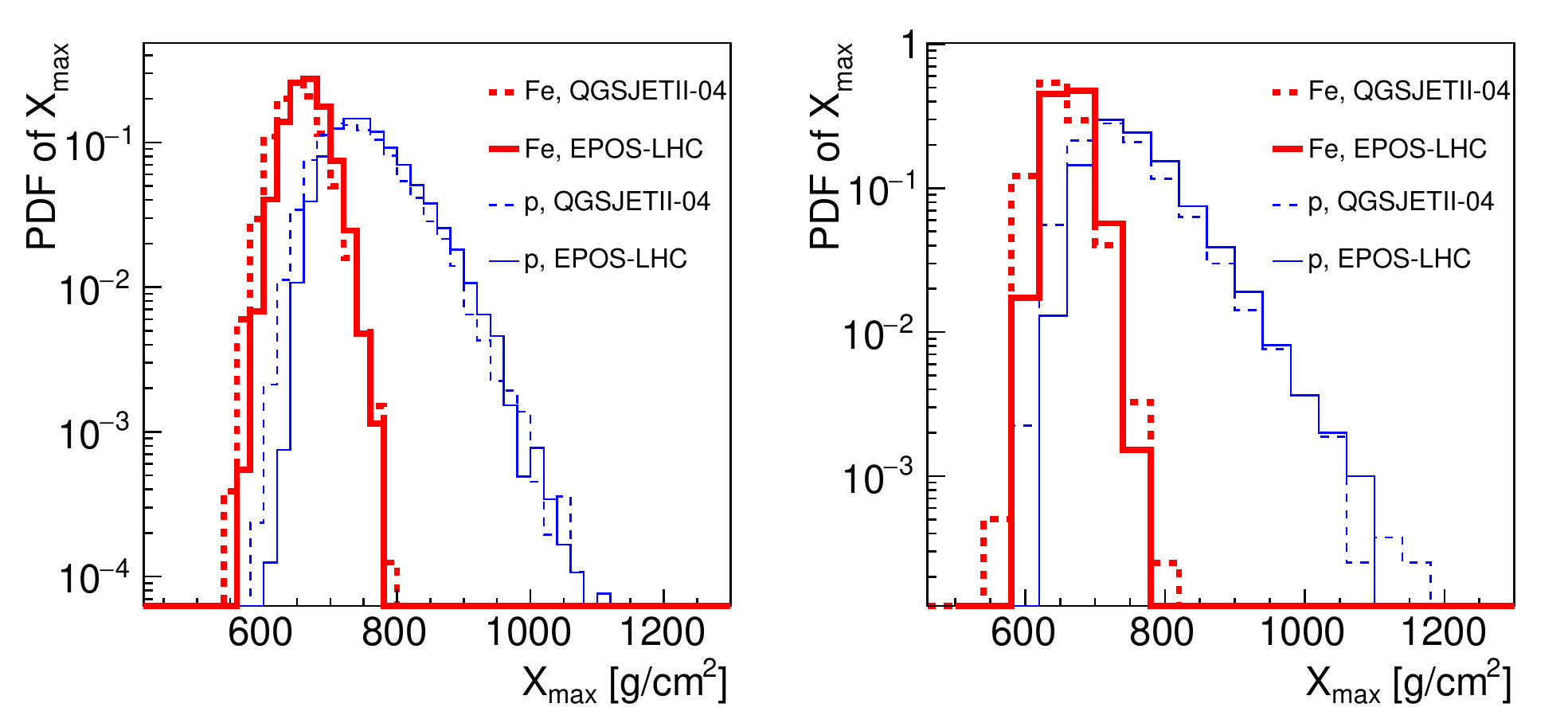}
    \caption{PDFs of $X_{\rm max}$ for proton and iron induced showers as predicted by QGSJETII-04 and EPOS-LHC hadronic interaction models in the energy interval $\lg E (\rm eV) = [18.2 - 18.3]$ for Auger \textit{(left)} and TA \textit{(right)} experiments. Note the difference in the bin width.}
    \label{fig_pdfs}
\end{figure}

An example of PDFs of $X_{\rm max}$ is given in Figure \ref{fig_pdfs} for proton and iron induced showers in the energy interval $\lg E (\rm eV)$ = [18.2--18.3] employing QGSJETII-04 and EPOS-LHC hadronic interaction models and taking into account the specific detector effects of \mbox{each experiment}. 

In the next Section we will use these PDFs to fit the experimental $X_{\rm max}$ distributions measured by Auger and TA experiments following a binned maximum-likelihood procedure.


\section{Fitting fractions of Auger and TA $X_{\rm max}$ distributions} \label{fit}

The experimental $X_{\rm max}$ distributions in each energy interval recorded by Auger and TA experiments are fitted with all possible combinations of primary elements (p, He, C, N, O, Ne, Si and Fe) following a binned maximum-likelihood procedure. Thus, we obtain the fractions of individual nuclei that best describe the observed $X_{\rm max}$ distributions. 
\begin{figure*}
    \includegraphics[width=\textwidth]{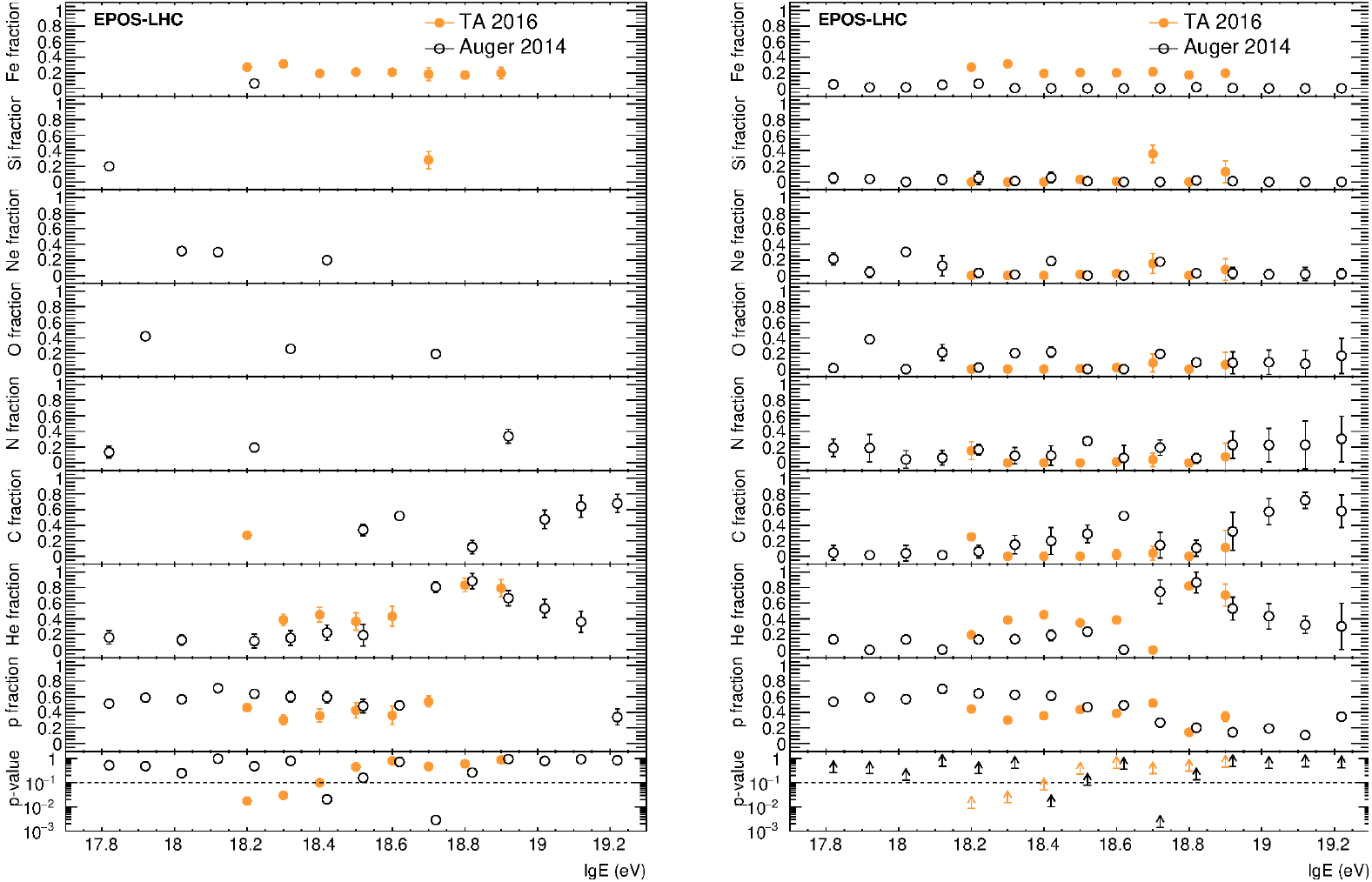}
    \caption{Fitted fractions of individual nuclei in each energy interval obtained with \textit{method} \#1 \textit{(left)} and \textit{method} \#2 \textit{(right)} employing the EPOS-LHC hadronic interaction model. Full orange circles stand for TA (2016) data while the black open circles represent the Auger (2014) data. The \textit{p-value} parameters are displayed on the bottom panels as one single value per energy interval \textit{(left)} and a range of \textit{p-values} represented with arrows \textit{(right)} (see text).} 
    \label{fig:2}
\end{figure*}
In this fitting procedure the minimizing quantity is $-\ln L$, which is defined as:
\begin{equation}
\label{logl}
 -\ln L = \sum_{i} y_i - n_i + n_i \ln(n_i / y_i) ,
\end{equation}
where $n_i$ represents the measured counts in the ``$i$''-th bin of the experimental $X_{max}$ distribution and $y_i$ is the MC prediction for that bin \cite{Baker:1983tu} given by $y_i = \sum_{j} [c_j f_j(i)]$ where $j = $ p, He, C, N, O, Ne, Si and Fe, $f_j(i)$ is an MC template and $c_j$ is the fraction of the $j$ component.

\begin{figure*}
    \includegraphics[width=\textwidth]{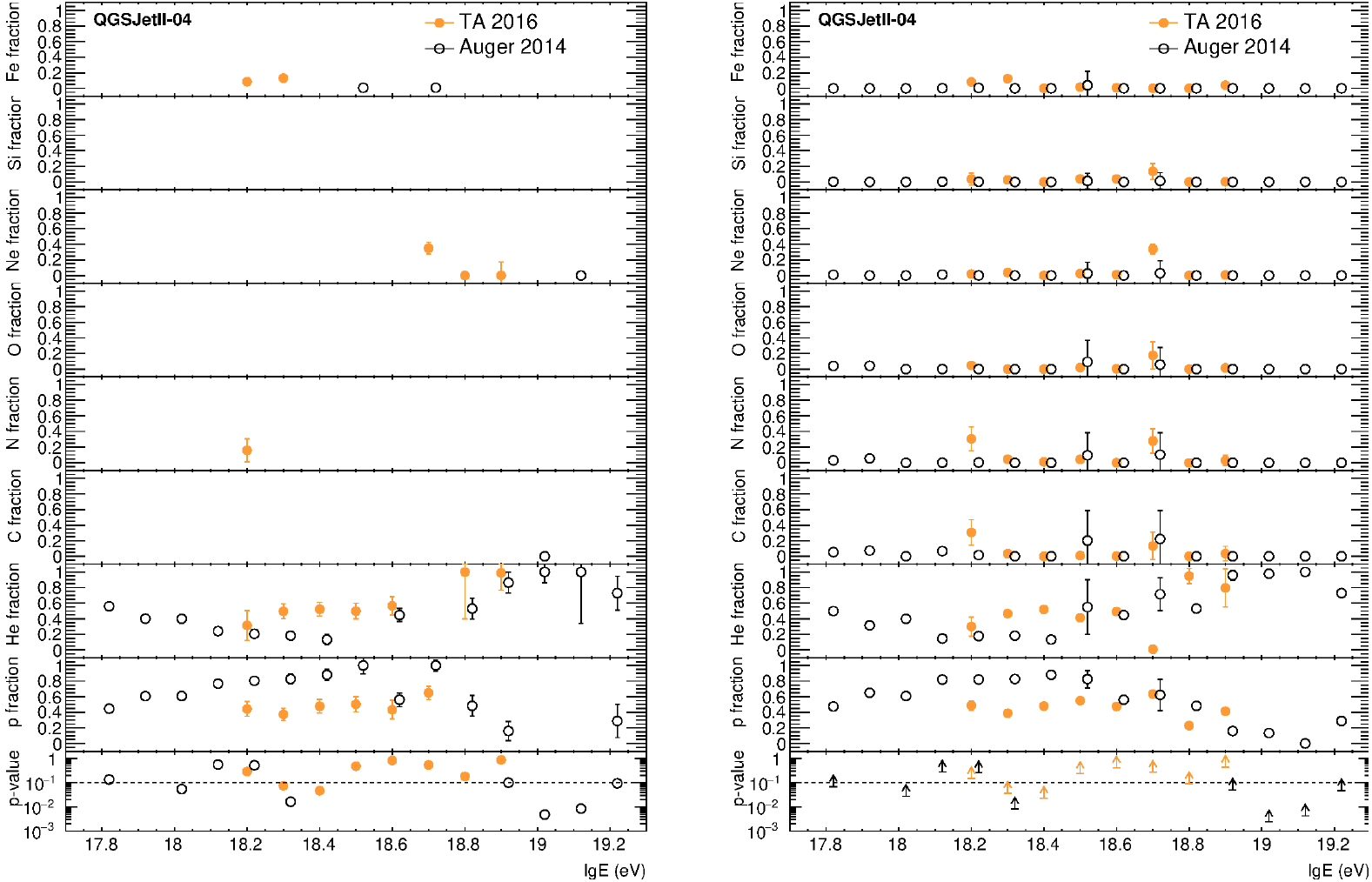}
    \caption{Fitted fractions of individual nuclei in each energy interval obtained with \textit{method} \#1 \textit{(left)} and \textit{method} \#2 \textit{(right)} employing the QGSJETII-04 hadronic interaction model. Full orange circles stand for TA (2016) data while the black open circles represent the Auger (2014) data. The \textit{p-value} parameters are displayed on the bottom panels as one single value per energy interval \textit{(left)} and a range of \textit{p-values} represented with arrows \textit{(right)} (see text).}     
    \label{fig:3}
\end{figure*}

\begin{figure*}
    \includegraphics[width=\textwidth]{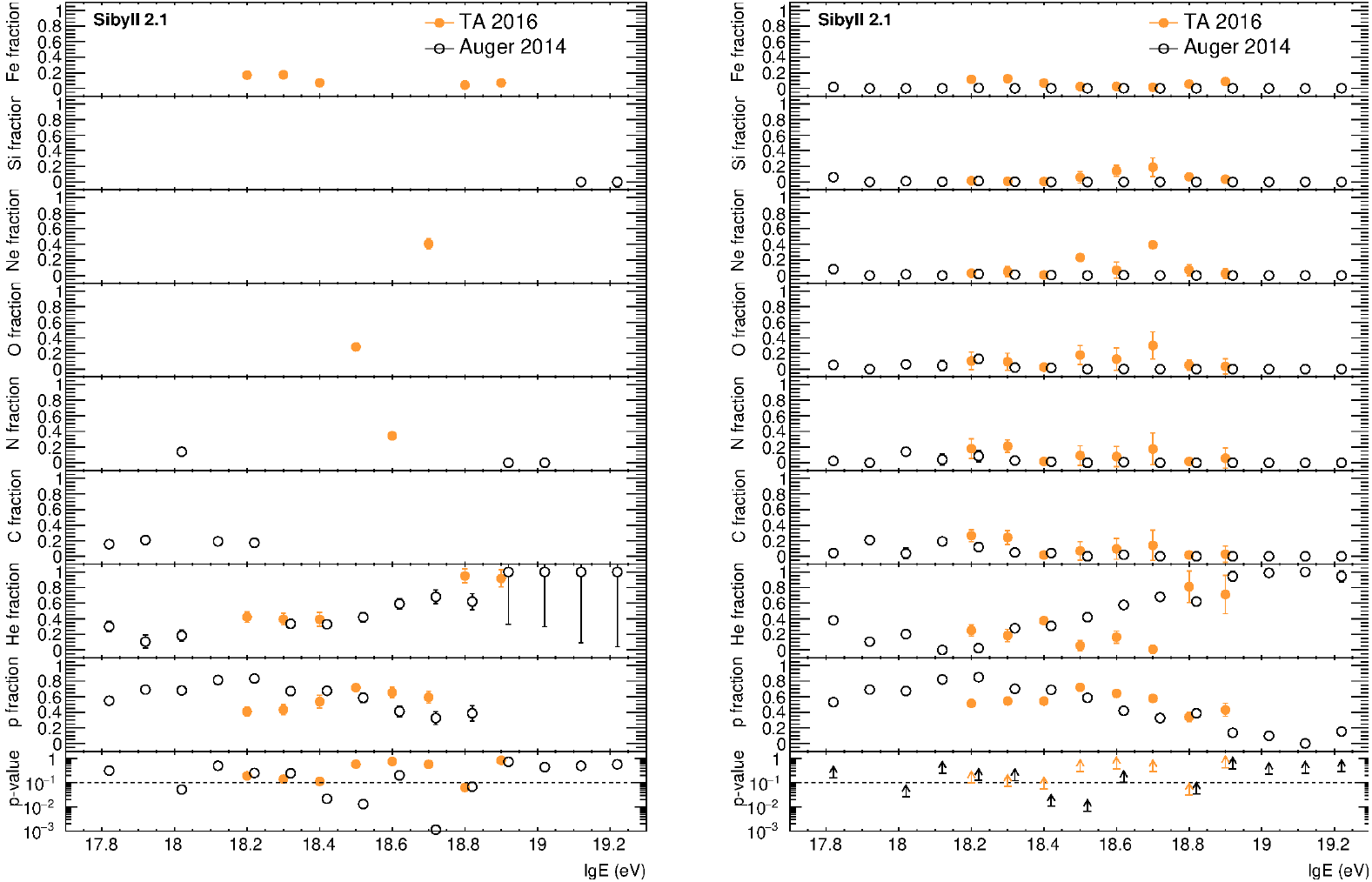}
    \caption{Fitted fractions of individual nuclei in each energy interval obtained with \textit{method} \#1 \textit{(left)} and \textit{method} \#2 \textit{(right)} employing the Sibyll 2.1 hadronic interaction model. Full orange circles stand for TA (2016) data while the black open circles represent the Auger (2014) data. The \textit{p-value} parameters are displayed on the bottom panels as one single value per energy interval \textit{(left)} and a range of \textit{p-values} represented with arrows \textit{(right)} (see text).}  
    \label{fig:4}
\end{figure*}

We quantify the goodness of fit using the \textit{p-value} parameter defined as:
\begin{equation}
\label{pval}
\textit{p-value} = 1 - \Gamma\left(\frac{ndf}{2}, \frac{\chi^2}{2}\right), 
\end{equation}
representing the probability of obtaining a worse fit than the one observed even if the distribution predicted by the fit results is correct. Here, $ndf$ represents the number of degrees of freedom computed as the number of bins (including empty bins) of each $X_{\rm max}$ distribution minus the number of elements considered in the fit, $\chi^2$ is the sum of the square of residuals using the parameters computed by the maximum-likelihood procedure and $\Gamma$ is the normalized upper incomplete Gamma function.
The results of the fitted fractions of $X_{\rm max}$ distributions measured by Auger and TA are presented following two methods.
In \textit{method} \#1 we obtained the ``best combination'' of elements from all possible combinations that best describe the experimental $X_{\rm max}$ distributions based on the highest \textit{p-value$^{max}$}, in each energy interval. The results are displayed in the left plots of \mbox{Figures \ref{fig:2}--\ref{fig:4}} for EPOS-LHC, QGSJETII-04 and Sibyll 2.1 hadronic interaction models. The error bars of each point were computed using the MINOS technique based on $\Delta L = 1 / 2$ rule \cite{eadie1971statistical}. 
The reconstructed fractions obtained for both experiments clearly show that the mass composition of primary UHECRs is dominated by light elements (p and He), which present a modulation of their abundances as a function of energy, but keeping the sum ($f_p + f_{He}$) roughly constant on the entire energy spectrum. This feature is predicted by all hadronic interaction models. An interesting aspect is the presence of Fe nuclei in a quite high abundance ($\sim$$20 \%$ predicted by EPOS-LHC model) on the entire energy spectrum of TA data which, at least qualitatively, seems to contradict the results of mass composition obtained on the basis of the first two moments of the $X_{\rm max}$ distributions \cite{Abbasi:2014sfa}.

\begin{figure}
  \begin{minipage}[b]{0.5\textwidth}
    \includegraphics[width=\textwidth]{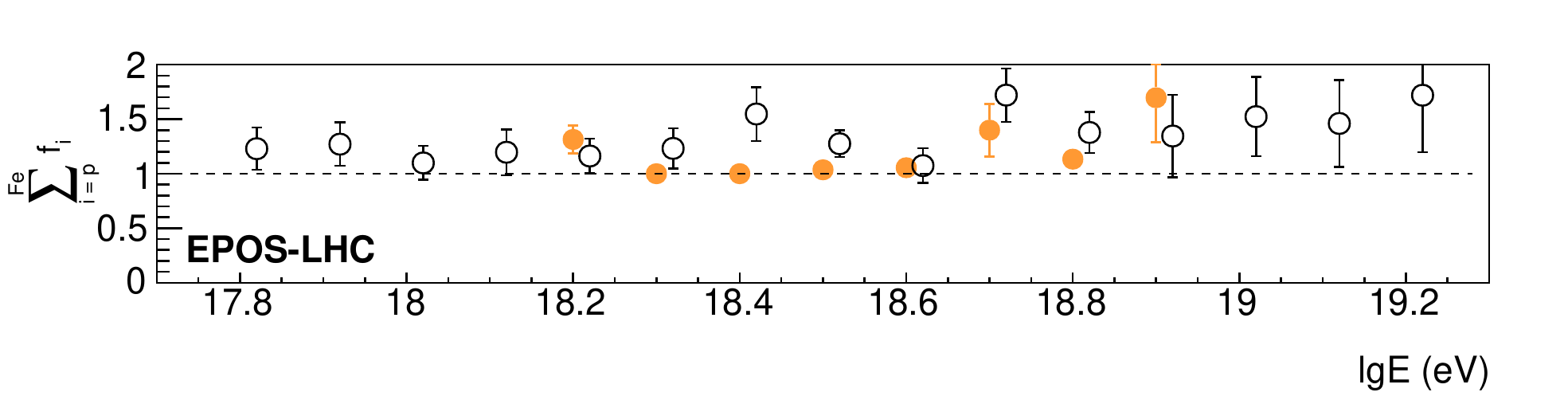}  
    \end{minipage}\par\vspace{-1.3\baselineskip}
    \begin{minipage}[b]{0.5\textwidth}
    \includegraphics[width=\textwidth]{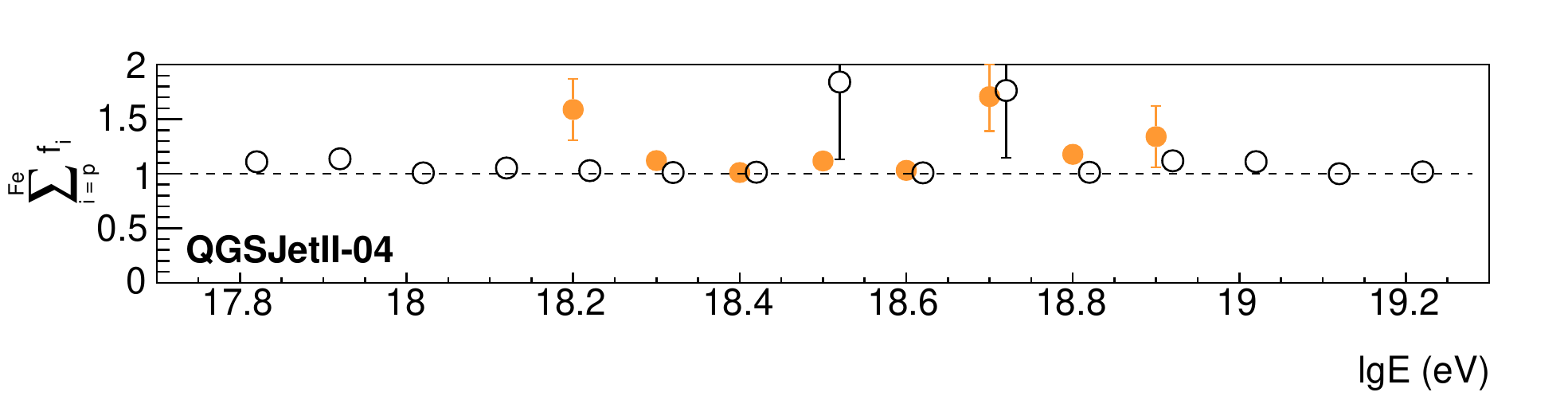}  
    \end{minipage}\par\vspace{-1.3\baselineskip}
    \begin{minipage}[b]{0.5\textwidth}
    \includegraphics[width=\textwidth]{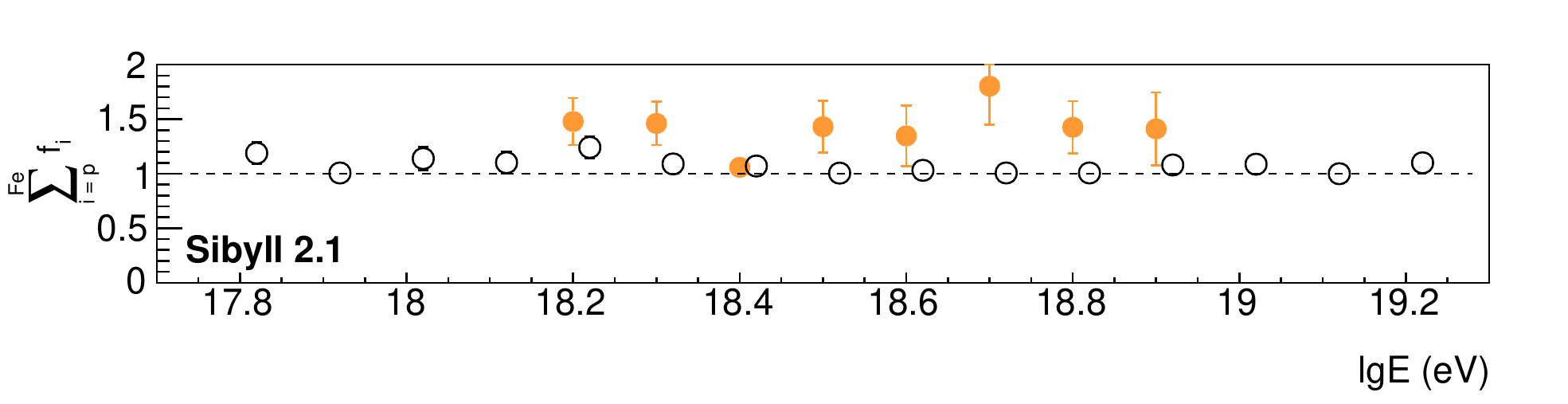}
    \end{minipage}
    \caption{The sum of all reconstructed fractions obtained with \textit{method} \#2 (see text) as a function of energy for EPOS-LHC \textit{(up)}, QGSJETII-04 \textit{(middle)} and Sibyll 2.1 \textit{(bottom)} hadronic interaction models.}
    \label{fig:7}
\end{figure}

In \textit{method} \#2 we extract the average of the reconstructed fractions of each species from all possible combinations of fitting elements whose goodness of fit parameter \textit{p-value} was in the range [0.5 $\cdot$ \textit{p-value$^{max}$}, \textit{p-value$^{max}$}]. For example, if the combination of the elements p + N + Fe and p + He + Si or simply p + He, and so on, give a good fit quality, then the average fraction of protons is the mean of the proton fractions reconstructed from all these combinations. In this case, the error bar associated to one species is computed as the square root of the sum of the squared uncertainties of the species obtained from different combinations of fitting elements whose goodness of fit was in the range mentioned above. In this way, we have an overview of the evolution of all fractions of individual nuclei as a function of primary energy. The results are displayed on the right plots of \mbox{Figures \ref{fig:2}--\ref{fig:4}} for the three hadronic interaction models. 
Both methods give similar results on the entire energy range.
It is worth mentioning that the shortcoming of \textit{method} \#2 is that the sum of all reconstructed fractions in a given energy interval will be biased (>1) as a consequence of the limitations of the fitting procedure when using extreme combinations of elements.
This bias is evaluated for each energy interval as the sum of all reconstructed fractions, $\sum_{i = p}^{Fe} f_i$, for all hadronic interaction models and is represented in Figure \ref{fig:7}.
As can be seen, the obtained values are not {too} far from $1$ for almost the entire energy spectrum of Auger data, while for TA we observe larger biases in many energy intervals, mainly due to the lower statistics in the experimental $X_{\rm max}$ distributions. Even so, we consider that these biases are not significant compared to the systematic and statistical uncertainties of the experimental $X_{\rm max}$ distributions.

In the next Section, we will perform an indirect comparison between the $X_{\rm max}$ distributions recorded by Auger and TA experiments using the individual fractions of nuclei reconstructed in this Section \ref{fit}.

\section{Auger vs. TA mass composition compatibility} \label{compatibility}

The measurements of $X_{\rm max}$ distributions obtained by both experiments cannot be directly compared because they use different approaches when analyzing the recorded events \cite{Yushkov:2019hoh}. 
This is the reason we produced two sets of MC predictions of PDFs of $X_{\rm max}$ taking into account the specific resolution and acceptance of each experiment. 

We test the compatibility of the measurements of $X_{\rm max}$ distributions recorded by both experiments following two approaches on the common energy range \mbox{$\lg E (\rm eV) =$ [18.2--19.0]}. 
In the \textit{first approach,} we translate the individual fractions of each primary species reconstructed from TA data in 
equivalent PDFs of $X_{\rm max}$ predicted for Auger (PDFs of $X_{\rm max}^{TA \rightarrow Auger} (i) = \sum_{j} c_{j}^{TA} f_{j}^{Auger} (i)$) and we perform the comparison PDFs of $X_{\rm max}^{TA \rightarrow Auger}$ vs. Auger data, while in the \textit{second approach} we translate the individual fractions of each primary species reconstructed from Auger data in equivalent PDFs of $X_{\rm max}$ predicted for TA (PDFs of $X_{\rm max}^{Auger \rightarrow TA} (i) = \sum_{j} c_{j}^{Auger} f_{j}^{TA} (i)$) and the comparison is performed between PDFs of $X_{\rm max}^{Auger \rightarrow TA}$ vs. TA data. 

In each approach, we use the fitting fractions obtained following \textit{method} \#1 and \#2 (see Section \ref{fit}).
Each comparison is performed for all hadronic interaction models and is characterized by three statistical tests: \textit{p-value} as goodness of fit, $KS$ and $AD$.
The \textit{p-value} parameter is obtained by fitting Auger data with PDFs of $X_{\rm max}^{TA \rightarrow Auger}$ and TA data with PDFs of $X_{\rm max}^{Auger \rightarrow TA}$ following the same binned maximum-likelihood procedure. 
Both $KS$ and $AD$ tests calculate the probability that two distributions come from the same parent distribution. 
The $KS$ test is one of the most used compatibility procedure with an improved response around the peak of the distributions. On the other hand, the $AD$ test is optimized to probe the compatibility with a higher accuracy on the tails of the distributions which, in our cases, are dominated by protons and He nuclei. 

\begin{figure*}
     \includegraphics[width=0.95\textwidth]{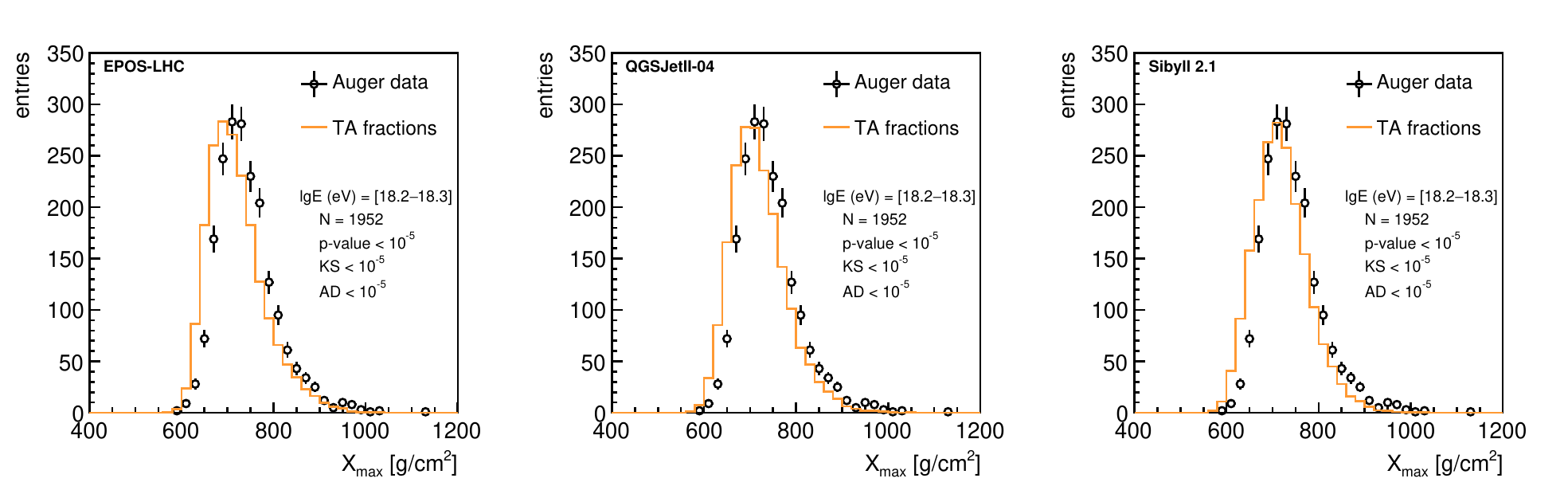}  
     \includegraphics[width=0.95\textwidth]{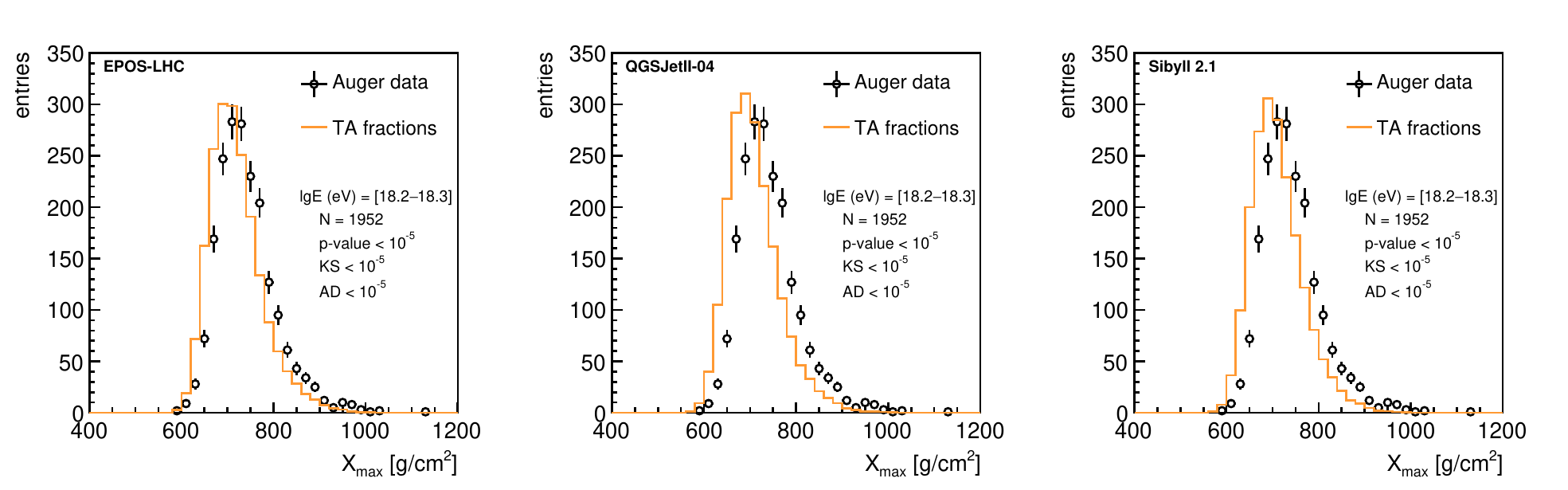} 
          \caption[width=0.9\textwidth]{The comparison between Auger data (black points) and PDFs of $X_{\rm max}^{TA \rightarrow Auger}$ in the energy interval $\lg E (\rm eV) = [18.2 - 18.3]$ using the TA fractions obtained with \textit{method} \#1 \textit{(top)} and \textit{method} \#2 \textit{(bottom)} for EPOS-LHC \textit{(left)}, QGSJETII-04 \textit{(middle)} and Sibyll 2.1 \textit{(right)} hadronic interaction models.}
     \label{fig:8}
     \includegraphics[width=0.95\textwidth]{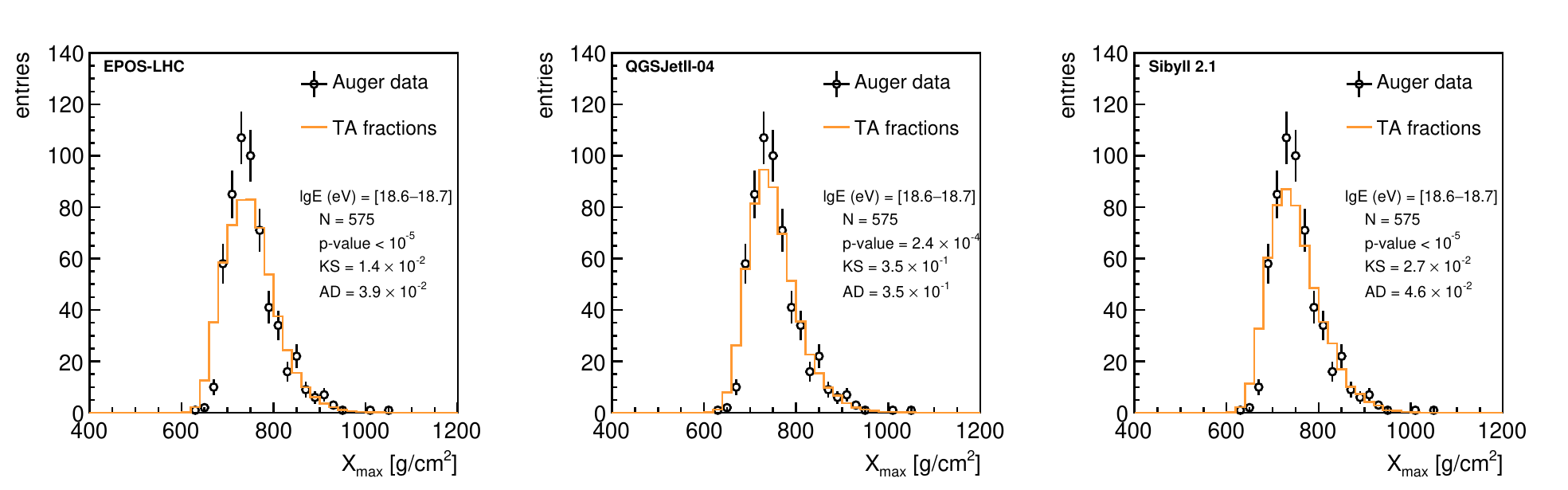}  
     \includegraphics[width=0.95\textwidth]{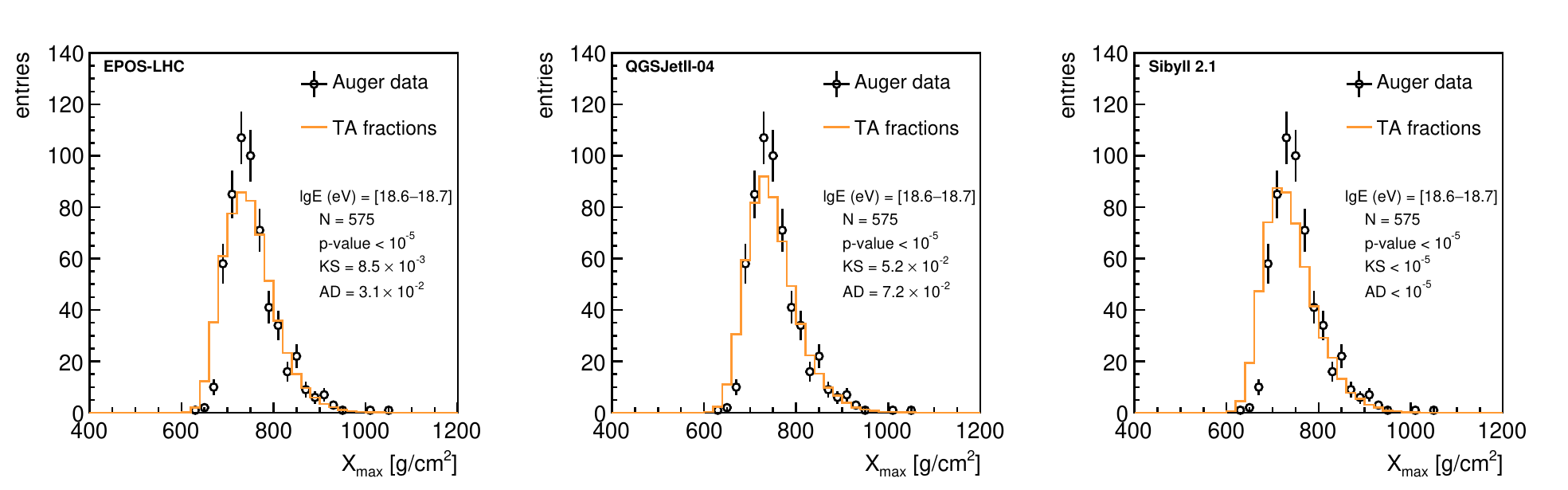}  
          \caption[width=0.9\textwidth]{The comparison between Auger data (black points) and PDFs of $X_{\rm max}^{TA \rightarrow Auger}$ in the energy interval $\lg E (\rm eV) = [18.6 - 18.7]$ using the TA fractions obtained with \textit{method} \#1 \textit{(top)} and \textit{method} \#2 \textit{(bottom)} for EPOS-LHC \textit{(left)}, QGSJETII-04 \textit{(middle)} and Sibyll 2.1 \textit{(right)} hadronic interaction models. }
     \label{fig:9}
 \end{figure*}

In Figure \ref{fig:8} we present the comparison between Auger data and PDFs of $X_{\rm max}^{TA \rightarrow Auger}$ for the energy interval $\lg E (\rm eV) =$ [18.2--18.3] considering the individual fitting fractions obtained from TA data by the two methods (see Section \ref{fit}) for all three hadronic interaction models. The statistics of the $X_{\rm max}$ distribution ($N = 1952$) as well as the \textit{p-value}, $KS$ and $AD$ parameters are displayed on the plots. We chose to give this example because it is the energy interval with the highest statistics of events recorded by both experiments.

 Figure \ref{fig:9} presents the same analysis as in Figure \ref{fig:8} but in the energy bin \mbox{$\lg E (\rm eV)$ = [18.6--18.7]} in which the number of events recorded by Auger is $N = 575$. In this case, the compatibility of the two datasets is much better, reaching $AD = 0.35$ for the QGSJetII-04 model. It is important to note that the highest degree of compatibility is obtained with the $AD$ test, which is more sensitive on the tails of the distributions dominated by protons and He nuclei. This is an indication that the measurements of both experiments are more compatible with respect to the light component (p and He) of primary UHECRs.

\begin{figure*}
     \includegraphics[width=0.95\textwidth]{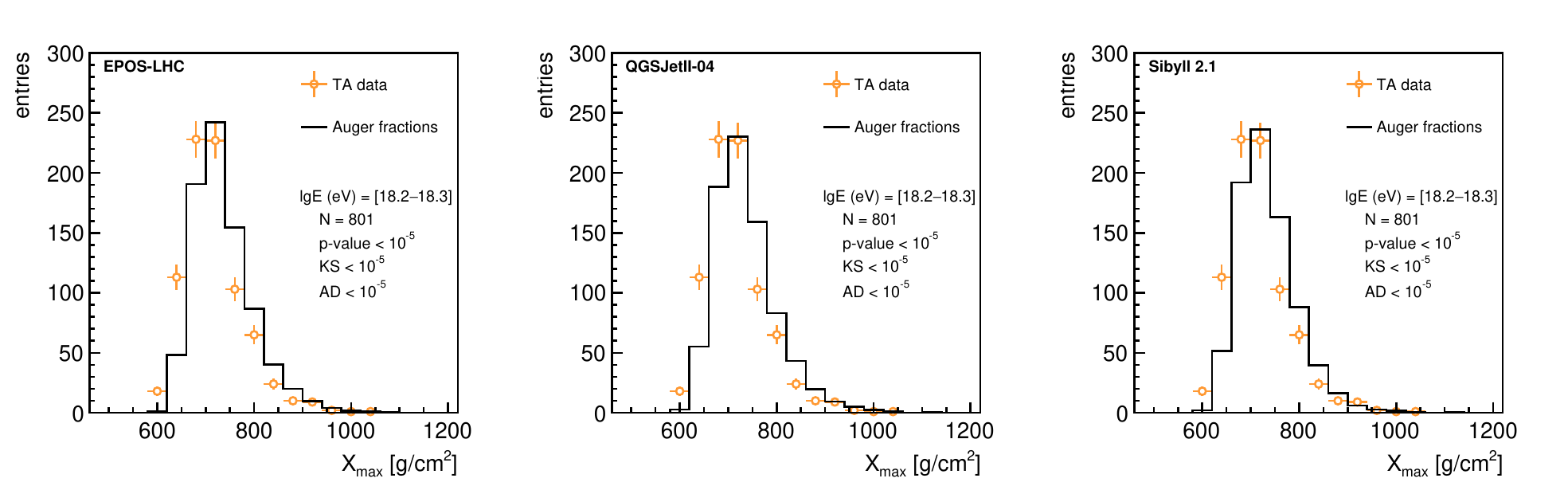}  
     \includegraphics[width=0.95\textwidth]{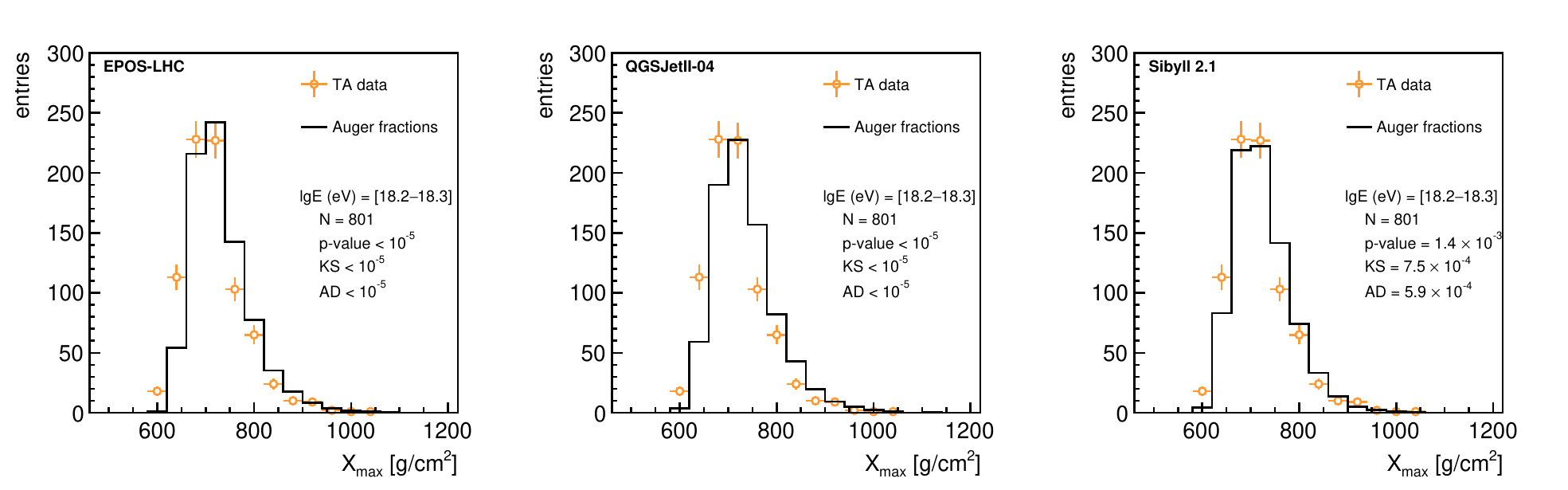}  
               \caption[width=0.9\textwidth]{The comparison between TA data (orange points) and PDFs of $X_{\rm max}^{Auger \rightarrow TA}$ in the energy interval $\lg E (\rm eV) = [18.2 - 18.3]$ using the Auger fractions obtained with \textit{method} \#1 \textit{(top)} and \textit{method} \#2 \textit{(bottom)} for EPOS-LHC \textit{(left)}, QGSJETII-04 \textit{(middle)} and Sibyll 2.1 \textit{(right)} hadronic interaction models.}
     \label{fig:10}
     \includegraphics[width=0.95\textwidth]{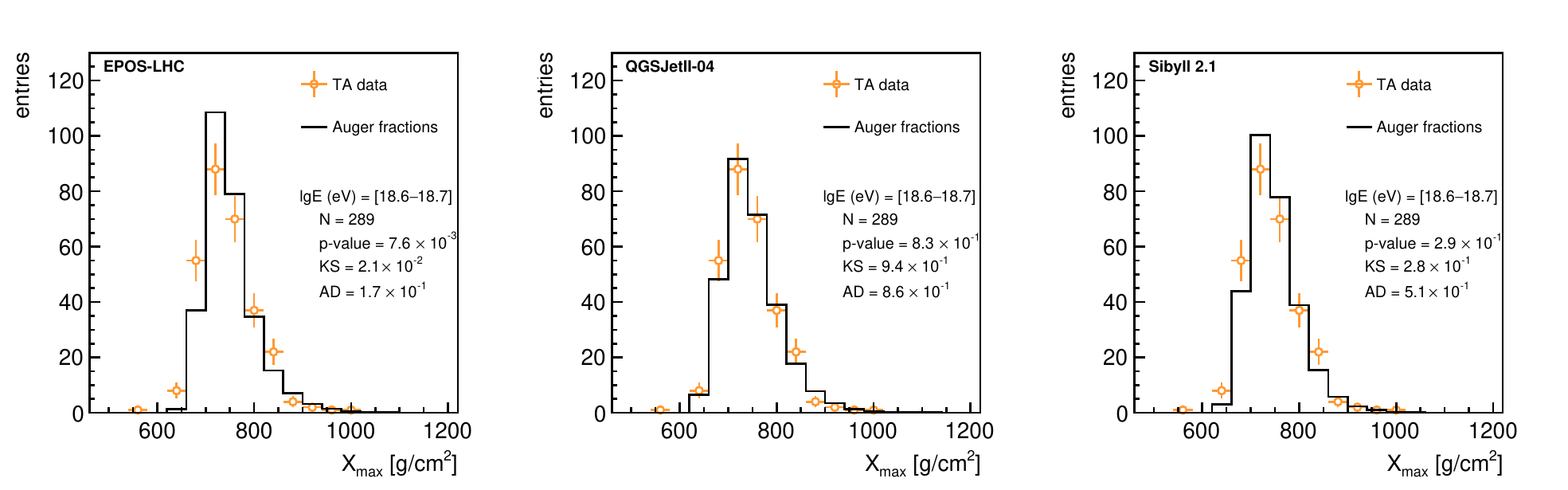}  
     \includegraphics[width=0.95\textwidth]{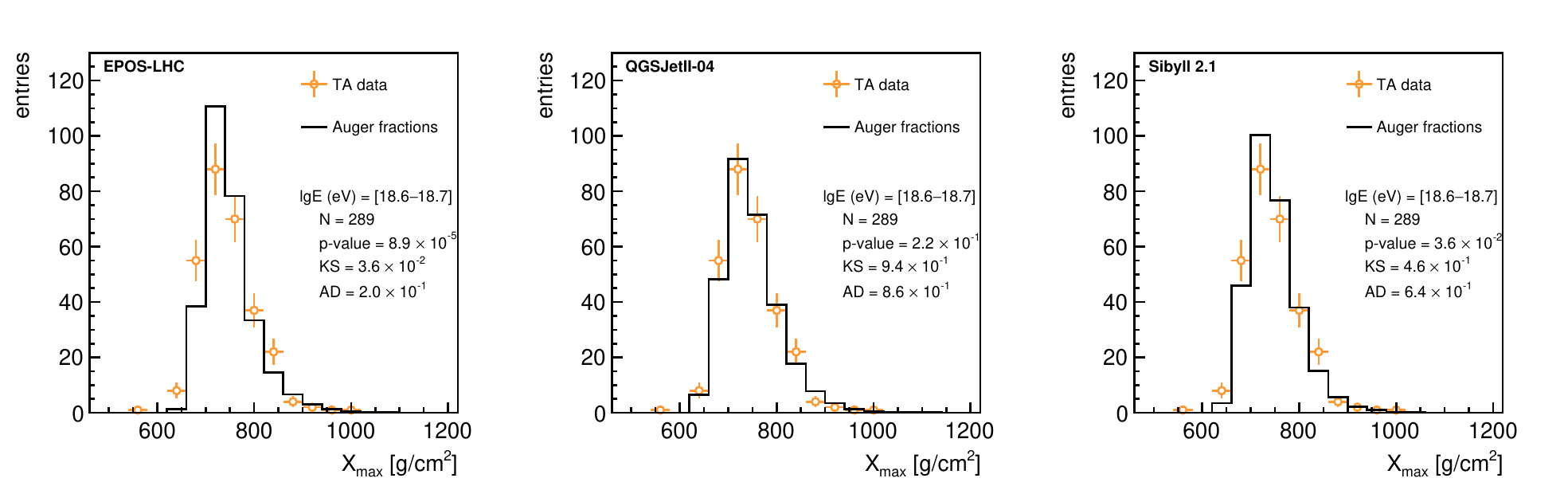}  
               \caption[width=0.9\textwidth]{The comparison between TA data (orange points) and PDFs of $X_{\rm max}^{Auger \rightarrow TA}$ in the energy interval $\lg E (\rm eV) = [18.6 - 18.7]$ using the Auger fractions obtained with \textit{method} \#1 \textit{(top)} and \textit{method} \#2 \textit{(bottom)} for EPOS-LHC \textit{(left)}, QGSJETII-04 \textit{(middle)} and Sibyll 2.1 \textit{(right)} hadronic interaction models.}
     \label{fig:11}
 \end{figure*}

In Figures \ref{fig:10} and \ref{fig:11} we present the comparison between TA data and the fitted fractions reconstructed from Auger data (\textit{second approach}) for the energy bins $\lg E (\rm eV)$ = [18.2--18.3] and $\lg E (\rm eV)$ = [18.6--18.7], respectively. 

As in the \textit{first approach,} the two experiments show that the degree of compatibility of the two datasets is good, reaching excellent agreement in some high energy intervals, especially when using the fractions reconstructed with \textit{method} \#2 (e.g., \textit{p-value}$ = 0.83$, $KS = 0.94$ and $AD = 0.86$) for $\lg E (\rm eV)$ = [18.6--18.7] considering QGSJetII-04 model. At lower energies, our study shows that the reconstruction methods need to be refined in order to achieve a better compatibility between the two sets of measurements.

The complete set of probabilities computed with the three statistical tests is listed in Tables \ref{tab1} and \ref{tab3} for the \textit{first approach} and \textit{second approach}, respectively.

\begin{table*}
\caption{The probabilities computed for the tree statistical tests of the comparison TA data vs. $X_{\rm max}^{Auger \rightarrow TA}$ considering the Auger fractions obtained using \textit{method} \#1 \textit{(left)} and \#2 \textit{(right)}.}
\label{tab3}
\begin{ruledtabular}
\begin{tabular}{l c c c c | c c c}

model & lg$(E/\rm eV)$ & $p-value$ & $KS$ & $AD$ & $p-value$ & $KS$ & $AD$\\
 
\colrule

& $[18.2 - 18.3]$ &	$< 10^{-5}$&	$< 10^{-5}$&	$< 10^{-5}$ &	$< 10^{-5}$ &	$< 10^{-5}$ &	$< 10^{-5}$\\
& $[18.3 - 18.4]$ &	$< 10^{-5}$&	$< 10^{-5}$&	$< 10^{-5}$ &	$< 10^{-5}$ &	$< 10^{-5}$ &	$< 10^{-5}$\\
& $[18.4 - 18.5]$ &	$2.0 \times 10^{-5}$&	$8.5 \times 10^{-5}$&	$1.3 \times 10^{-3}$ &	$< 10^{-5}$ &	$7.5 \times 10^{-2}$ &	$4.2 \times 10^{-2}$\\
EPOS-LHC& $[18.5 - 18.6]$ &	$2.7 \times 10^{-4}$&	$2.1 \times 10^{-3}$&	$4.4 \times 10^{-2}$ &	$2.8 \times 10^{-5}$ &	$8.0 \times 10^{-2}$ &	$4.8 \times 10^{-2}$ \\
& $[18.6 - 18.7]$ &	$7.6 \times 10^{-3}$&	$2.2 \times 10^{-2}$&	$1.7 \times 10^{-1}$ &	$9.0 \times 10^{-5}$ &	$3.6 \times 10^{-2}$ &	$1.9 \times 10^{-1}$\\
& $[18.7 - 18.8]$ &	$2.1 \times 10^{-5}$&	$1.8 \times 10^{-2}$&	$1.1 \times 10^{-2}$ &	$< 10^{-5}$ &	$7.5 \times 10^{-2}$ &	$2.5 \times 10^{-2}$\\
& $[18.8 - 18.9]$ &	$8.5 \times 10^{-2}$&	$1.5 \times 10^{-1}$&	$4.4 \times 10^{-1}$ &	$8.5 \times 10^{-4}$ &	$2.7 \times 10^{-1}$ &	$4.8 \times 10^{-1}$\\
& $[18.9 - 19.0]$ &	$6.3 \times 10^{-1}$&	$9.0 \times 10^{-1}$&	$7.7 \times 10^{-1}$ &	$8.2 \times 10^{-2}$ &	$9.6 \times 10^{-1}$ &	$7.6 \times 10^{-1}$\\

\colrule
& $[18.2 - 18.3]$ &	$< 10^{-5}$&	$< 10^{-5}$&	$< 10^{-5}$ &	$< 10^{-5}$ &	$< 10^{-5}$ &	$< 10^{-5}$\\ 
& $[18.3 - 18.4]$ &	$< 10^{-5}$&	$< 10^{-5}$&	$< 10^{-5}$ &	$< 10^{-5}$ &	$< 10^{-5}$ &	$< 10^{-5}$\\
& $[18.4 - 18.5]$ &	$< 10^{-5}$&	$3.2 \times 10^{-2}$&	$4.3 \times 10^{-3}$ &	$< 10^{-5}$ &	$3.2 \times 10^{-2}$ &	$4.3 \times 10^{-3}$ \\
QGSJETII-04& $[18.5 - 18.6]$ &	$4.0 \times 10^{-5}$&	$4.4 \times 10^{-3}$&	$2.0 \times 10^{-3}$ &	$< 10^{-5}$ &	$4.7 \times 10^{-2}$ &	$6.7 \times 10^{-3}$\\
& $[18.6 - 18.7]$ &	$8.3 \times 10^{-1}$&	$9.4 \times 10^{-1}$&	$8.6 \times 10^{-1}$ &	$2.2 \times 10^{-1}$ &	$9.4 \times 10^{-1}$ &	$8.6 \times 10^{-1}$\\
& $[18.7 - 18.8]$ &	$4.4 \times 10^{-5}$&	$< 10^{-5}$&	$5.7 \times 10^{-4}$ &	$4.0 \times 10^{-4}$ &	$5.6 \times 10^{-1}$ &	$5.5 \times 10^{-1}$\\
& $[18.8 - 18.9]$ &	$7.9 \times 10^{-2}$&	$7.4 \times 10^{-1}$&	$3.7 \times 10^{-1}$ &	$8.7 \times 10^{-4}$ &	$7.4 \times 10^{-1}$ &	$3.7 \times 10^{-1}$\\
& $[18.9 - 19.0]$ &	$9.0 \times 10^{-1}$&	$1.0$&	$1.0$ &	$2.5 \times 10^{-1}$ &	$1.0$ &	$1.0$ \\

\colrule
& $[18.2 - 18.3]$ &	$< 10^{-5}$&	$< 10^{-5}$&	$< 10^{-5}$ &	$1.4 \times 10^{-3}$ &	$7.5 \times 10^{-4}$ &	$3.0 \times 10^{-4}$\\
& $[18.3 - 18.4]$ &	$< 10^{-5}$&	$< 10^{-5}$&	$< 10^{-5}$ &	$< 10^{-5}$ &	$< 10^{-5}$ &	$< 10^{-5}$\\
& $[18.4 - 18.5]$ &	$3.5 \times 10^{-3}$&	$3.9 \times 10^{-3}$&	$1.9 \times 10^{-2}$ &	$1.5 \times 10^{-3}$ &	$9.0 \times 10^{-2}$ &	$1.6 \times 10^{-1}$ \\
Sibyll 2.1& $[18.5 - 18.6]$ &	$4.5 \times 10^{-2}$&	$4.9 \times 10^{-2}$&	$1.5 \times 10^{-1}$ &	$4.5 \times 10^{-2}$ &	$4.9 \times 10^{-2}$ &	$1.5 \times 10^{-1}$ \\
& $[18.6 - 18.7]$ &	$2.9 \times 10^{-1}$&	$2.8 \times 10^{-1}$&	$5.1 \times 10^{-1}$ &	$3.6 \times 10^{-2}$ &	$4.6 \times 10^{-1}$ &	$6.4 \times 10^{-1}$\\
& $[18.7 - 18.8]$ &	$2.8 \times 10^{-3}$&	$9.9 \times 10^{-2}$&	$7.9 \times 10^{-2}$ &	$1.5 \times 10^{-5}$ &	$9.9 \times 10^{-2}$ &	$7.9 \times 10^{-2}$\\
& $[18.8 - 18.9]$ &	$4.8 \times 10^{-2}$&	$4.2 \times 10^{-1}$&	$4.5 \times 10^{-1}$ &	$4.0 \times 10^{-4}$ &	$4.2 \times 10^{-1}$ &	$4.5 \times 10^{-1}$\\
& $[18.9 - 19.0]$ &	$8.2 \times 10^{-1}$&	$9.9 \times 10^{-1}$&	$7.7 \times 10^{-1}$ &	$1.9 \times 10^{-1}$ &	$1.0$    & 	$1.0$ 

\end{tabular}
\end{ruledtabular}
\end{table*}

\begin{table*}
\caption{The probabilities computed for the tree statistical tests of the comparison Auger data vs. $X_{\rm max}^{TA \rightarrow Auger}$ considering the TA fractions obtained using \textit{method} \#1 \textit{(left)} and \#2 \textit{(right)}.}
\label{tab1}
\begin{ruledtabular}
\begin{tabular}{l c c c c | c c c}

model & lg$(E/\rm eV)$ & $p-value$ & $KS$ & $AD$ & $p-value$ & $KS$ & $AD$\\
\colrule

&$[18.2 - 18.3]$&	$< 10^{-5}$&   	$< 10^{-5}$&    	$< 10^{-5}$ &	$< 10^{-5}$ &	$< 10^{-5}$ &	$< 10^{-5}$\\
&$[18.3 - 18.4]$&	$< 10^{-5}$&	    $< 10^{-5}$&   	$< 10^{-5}$ &	$< 10^{-5}$ &	$< 10^{-5}$ &	$< 10^{-5}$\\
&$[18.4 - 18.5]$&	$< 10^{-5}$&   	$< 10^{-5}$&    	$< 10^{-5}$ &	$< 10^{-5}$ &	$< 10^{-5}$ &	$< 10^{-5}$\\
EPOS-LHC&$[18.5 - 18.6]$&	$< 10^{-5}$&   	$< 10^{-5}$&    	$1.0 \times 10^{-3}$ &	$< 10^{-5}$ &	$< 10^{-5}$  &	$1.9 \times 10^{-4}$\\
&$[18.6 - 18.7]$&	$< 10^{-5}$&   	$1.4 \times 10^{-2}$&    	$3.9 \times 10^{-2}$ &	$< 10^{-5}$ &	$8.4 \times 10^{-3}$ &	$3.1 \times 10^{-2}$\\
&$[18.7 - 18.8]$&	$< 10^{-5}$&   	$< 10^{-5}$&    	$8.5 \times 10^{-5}$ &	$< 10^{-5}$ &	$< 10^{-5}$ &	$< 10^{-5}$\\
&$[18.8 - 18.9]$&	$1.1 \times 10^{-2}$&   	$7.7 \times 10^{-3}$&    	$2.4 \times 10^{-2}$ &	$1.4 \times 10^{-4}$ &	$6.8 \times 10^{-2}$ &	$1.3 \times 10^{-1}$\\
&$[18.9 - 19.0]$&	$3.8 \times 10^{-1}$&   	$5.2 \times 10^{-2}$&    	$2.5 \times 10^{-1}$ &	$5.2 \times 10^{-4}$ &	$5.4 \times 10^{-2}$ &	$1.9 \times 10^{-1}$ \\

\colrule

&$[18.2 - 18.3]$&	$< 10^{-5}$&   	$< 10^{-5}$&    	$< 10^{-5}$ &	$< 10^{-5}$ &	$< 10^{-5}$ &	$< 10^{-5}$\\
&$[18.3 - 18.4]$&	$< 10^{-5}$&   	$< 10^{-5}$&    	$< 10^{-5}$ &	$< 10^{-5}$ &	$< 10^{-5}$ &	$< 10^{-5}$\\
&$[18.4 - 18.5]$&	$< 10^{-5}$&   	$2.1 \times 10^{-4}$&    	$3.6 \times 10^{-5}$ &	$< 10^{-5}$ &	$8.2 \times 10^{-5}$ &	$1.2 \times 10^{-5}$\\
QGSJETII-04&$[18.5 - 18.6]$&	$< 10^{-5}$&   	$1.1 \times 10^{-2}$&    	$2.5 \times 10^{-2}$ &	$< 10^{-5}$ &	$< 10^{-5}$ &	$< 10^{-5}$\\
&$[18.6 - 18.7]$&	$2.5 \times 10^{-4}$&   	$3.5 \times 10^{-1}$&    	$3.6 \times 10^{-1}$ &	$< 10^{-5}$ &	$5.2 \times 10^{-2}$ &	$7.2 \times 10^{-2}$\\
&$[18.7 - 18.8]$&	$< 10^{-5}$&   	$6.1 \times 10^{-5}$&    	$6.3 \times 10^{-4}$ &	$< 10^{-5}$ &	$< 10^{-5}$ &	$< 10^{-5}$\\
&$[18.8 - 18.9]$&	$< 10^{-5}$&   	$< 10^{-5}$&    	$2.1 \times 10^{-4}$ &	$< 10^{-5}$ &	$6.1 \times 10^{-4}$ &	$8.9 \times 10^{-3}$\\
&$[18.9 - 19.0]$&	$7.9 \times 10^{-2}$&   	$1.6 \times 10^{-2}$&    	$8.1 \times 10^{-2}$ &	$4.1 \times 10^{-4}$ &	$1.7 \times 10^{-2}$ & $8.6 \times 10^{-2}$\\

\colrule

&$[18.2 - 18.3]$&	$< 10^{-5}$&   	$< 10^{-5}$&    	$< 10^{-5}$ &	$< 10^{-5}$ &	$< 10^{-5}$ &	$< 10^{-5}$\\
&$[18.3 - 18.4]$&	$< 10^{-5}$&   	$< 10^{-5}$&    	$< 10^{-5}$ &	$< 10^{-5}$ &	$< 10^{-5}$ &	$< 10^{-5}$\\
&$[18.4 - 18.5]$&	$< 10^{-5}$&   	$8.4 \times 10^{-4}$&    	$1.4 \times 10^{-4}$ &	$< 10^{-5}$ &	$< 10^{-5}$ &	$< 10^{-5}$\\
Sibyll 2.1&$[18.5 - 18.6]$&	$< 10^{-5}$&   	$6.8 \times 10^{-5}$&    	$1.6 \times 10^{-3}$ &	$< 10^{-5}$ &	$< 10^{-5}$ &	$< 10^{-5}$\\
&$[18.6 - 18.7]$&	$< 10^{-5}$&   	$2.7 \times 10^{-2}$&    	$4.7 \times 10^{-2}$ &	$< 10^{-5}$ &	$< 10^{-5}$ &	$< 10^{-5}$\\
&$[18.7 - 18.8]$&	$< 10^{-5}$&   	$5.5 \times 10^{-5}$&    	$5.9 \times 10^{-4}$ &	$< 10^{-5}$ &	$< 10^{-5}$ &	$< 10^{-5}$\\
&$[18.8 - 18.9]$&	$5.4 \times 10^{-5}$&   	$< 10^{-5}$&    	$1.8 \times 10^{-4}$ &	$< 10^{-5}$ &	$< 10^{-5}$ &	$2.4 \times 10^{-4}$\\
&$[18.9 - 19.0]$&	$3.8 \times 10^{-1}$&   	$1.5 \times 10^{-2}$&    	$2.9 \times 10^{-2}$ &	$1.7 \times 10^{-3}$ &	$1.4 \times 10^{-2}$ & $7.9 \times 10^{-2}$

\end{tabular}
\end{ruledtabular}
\end{table*}


{ \section{Discussions and Conclusions}} \label{conclusions}

In this paper, we present an alternative approach to inferring the mass composition of the primary UHECRs using the available $X_{\rm max}$ distributions recorded by Auger (2014) and TA (2016) experiments, by comparisons with MC templates predicted by EPOS-LHC, QGSJETII-04 and Sibyll 2.1 hadronic interaction models.

A common general remark is that the measurements of both experiments suggest that the mass composition of primary UHECRs is dominated ($\gtrsim$$70\%$) by protons and He nuclei, which present a modulation of their abundances as a function of energy but keeping the sum ($f_p + f_{He}$) roughly constant on the entire energy spectrum. This conclusion holds for all the three hadronic interaction models. An interesting aspect is the presence of Fe nuclei in a quite high abundance ($\sim$$20 \%$) in TA data on the entire energy spectrum \mbox{$\lg E (\rm eV) =$ [18.2--19.0]} as predicted by the EPOS-LHC hadronic interaction model.

The general conclusion is that, using the approaches proposed in this paper, the degree of compatibility of the two datasets is good, reaching excellent agreement in some high energy intervals above the \textit{ankle}, as computed by the three statistical tests (e.g., \textit{p-value}$ = 0.83$, $KS = 0.94$ and $AD = 0.86$ for $\lg E (\rm eV) =$ [18.6--18.7] considering the QGSJetII-04 model). The best agreement is achieved when the indirect comparison is performed using the reconstructed fractions of nuclei from Auger data to build the equivalent distributions, PDFs of $X_{\rm max}^{Auger \rightarrow TA}$, which are compared with TA data. However, our study reveals that, at low energies, further effort in terms of data analysis is required in order to harmonize the results of the two experiments, taking into account that the models are not able to accurately describe the data in some energy intervals.

We believe that the current approach could be very useful in future studies on mass composition, especially to crosscheck the measurements of $X_{\rm max}$ between Auger and TA once the statistics of the recorded events will increase.

\subsection*{Acknowledgments}
I would like to thank Octavian Sima and Ionel Lazanu for many useful discussions and suggestions.
This work was supported by a grant of the Romanian Ministry of Education and Research, CNCS - UEFISCDI, project number PN-III-P1-1.1-PD-2019-0178, within PNCDI III.

\bibliography{Auger_TA_fractions}

\end{document}